# Memory effect by coupling between translational and rotational Brownian motion in water–ethanol mixtures

Ken Judai[1]*, Satoshi Shibuta[1,2], and Kazuki Furukawa[1,3]

[1]Department of Physics, College of Humanities and Sciences, Nihon University, Sakurajosui 3-25-40, Setagaya-ku, Tokyo 156-8550, Japan

[2]National Institute of Technology, Kagoshima College, Shinko Hayato 1460-1, Kirishima, Kagoshima 899-5193, Japan

[3]National Institute of Technology, Ariake College, Higashihagio-Machi 150, Omuta, Fukuoka 836-8585, Japan

ABSTRACT:

The Brownian motion in water–ethanol mixtures exhibits abnormally large displacements. Using falling-ball viscometry applied to colloidal particles, we experimentally verified that no anomaly exists in the viscosity coefficient of the solution. Our findings reveal that the anomalous Brownian motion can be attributed to the memory effect arising from the coupling between rotational and translational motion. It was deduced that the relaxation time of sub-microsecond rotational motion serves as a memory effect, influencing translational motion and leading to anomalously large Brownian displacements over a timescale of seconds. This study demonstrates that short-time memory effects can be effectively investigated through long-duration measurements.

judai.ken@nihon-u.ac.jp

*author to whom correspondence should be addressed

## I. INTRODUCTION

In the early 19th century, botanist Robert Brown systematically studied the erratic motion of microparticles suspended in a liquid [1]. Under a microscope, the motion of a microparticle was observed to be entirely independent of its previous step, indicating that a defining characteristic of Brownian motion is its inherent randomness. The origin of this random motion was first formulated by Einstein, who incorporated the collisions of surrounding fluid molecules into the framework of statistical mechanics [2]. Since then, the theory of Brownian motion, as a theoretical model of accumulated small random changes, has profoundly influenced fields such as nonequilibrium and nonlinear statistical mechanics, noise theory [3], financial engineering [4], and numerous other disciplines.

Einstein's formulation also enabled proof of the existence of atoms and molecules by connecting fluid mechanics and statistical mechanics. Perrin experimentally observed the motion of colloidal particles and provided empirical validation of atomic and molecular existence [5]. Specifically, he measured the mean-square displacement (MSD), denoted as $\langle \Delta x^2 \rangle$, of particles as a function of the time interval $t$:

$$\langle \Delta x^2 \rangle = 2Dt = 2 \cdot \frac{k_B T}{6\pi \eta a} \cdot t, \qquad (1)$$

where $k_B$ denotes the Boltzmann constant, $T$ denotes the temperature, $\eta$ is the viscosity, and $a$ is the radius of the particle. The diffusion coefficient $D$, which quantifies particle mobility, is determined by the ratio of thermal fluctuation magnitude to the viscous resistance experienced by the particles. Notably, the Stokes resistance term ($\gamma = 6\pi \eta a$) accounts for the hydrodynamic radius of the particles, which may need correction in cases where solvent molecules co-move with the particles.

Mathematically, Langevin derived equations describing the dynamics of random processes [6], paving the way for advanced analyses. As with Einstein's approach, Langevin's model assumes that molecular collisions relax in a short time, with viscous drag dominating on the timescales of observation. In this context, it is assumed that the observed timescales are significantly longer than the characteristic time $\tau_p = m/\gamma$, where $m$ is the mass of the particle. However, at shorter timescales, the MSD deviates from being proportional to time, instead varying with the square of time, a phenomenon known as ballistic diffusion [7].

The Langevin equation has also been extended to account for memory effects [8,9]. When a particle moves, the surrounding solvent flows into the space it leaves behind, creating a memory effect where the solvent retains traces of past particle motion and exerts a drag. In typical experimental setups, this memory effect occurs too quickly to be observed, allowing Brownian motion to be treated as a memoryless process. However, photo-trap experiments and ultra-fast observations have successfully confirmed the existence of memory effects [10-12]. While Brownian motion remains a fundamental physical phenomenon, much of the ongoing research focuses on scenarios beyond ultra-short timescales, where memory effects are negligible. Active Brownian motion has been explored using Janus particles [13,14], and dynamic light scattering methods [15,16] have been widely employed to measure nanoparticle size distributions [17–19].

Our research group has reported deviations from conventional Brownian behavior in water–alcohol mixtures [20–22]. The MSD in these mixtures, including mono-alcohols such as methanol, ethanol, 1-propanol [21], and $t$-butyl alcohol [22], exhibits significantly larger displacements than predicted by the Stokes–Einstein equation (Eq. (1)) when using standard viscosity values. This anomalously large displacement cannot be explained by simple viscosity corrections or changes in the hydrodynamic radius, as increased molecular solvation typically reduces particle displacement. Consequently, a hypothesis was proposed suggesting a coupling between the translational and rotational energies of the particles [20,23].

Due to the limited molecular-level understanding of water–alcohol mixtures, numerous experimental [24–33] and theoretical studies [34–38] have been conducted using small-angle diffraction methods [24–26], spectroscopic techniques [27–31], thermodynamic measurements [32,33], and molecular dynamics (MD) simulations [34–36]. While experimental studies provide static information, dynamic insights intrinsic to Brownian motion in molecular mixtures can unveil entirely new molecular phenomena.

In this study, we report a memory effect observed in colloidal particles suspended in a water–ethanol mixture. A falling-ball experiment [39] was performed to measure the viscosity of colloidal particles under the influence of gravity. This approach separates the one-dimensional falling velocity due to gravitational force from the multidirectional Brownian motion. Through these measurements, we precisely determined the Stokes drag term in Eq. (1), which depends on both the viscous coefficient and the hydrodynamic radius of the particle. Our results revealed no discrepancy between the measured viscosity and values reported in the literature. Further analysis demonstrated that thermal disturbances are responsible for the anomalous Brownian motion, and the coupling between rotational and translational motion leads to a greater MSD due to memory effects.

**II. EXPERIMENT**

Video microscopy and particle-tracking or particle-counting techniques have advanced rapidly in recent years [40]. In our experiments, we utilized commercially available monodispersed polystyrene beads (1.0 μm Polybead® Microspheres, Polysciences, Inc.). Ultrapure water (Simplicity® UV, Millipore SAS) and ethanol (>99.5%, FUJIFILM Wako Pure Chemical Corp.) were mixed at varying weight concentrations. Polystyrene beads were suspended in the water–ethanol mixtures at a fixed concentration (400-fold dilution of a 2.5 wt% commercial suspension). The prepared liquid was introduced into a cylindrical hole (1.1 mm in height and 1.0 mm in diameter) within a thermally stabilized aluminum block. Both ends of the hole were sealed with glass coverslips using silicon grease (BARRIERTA, Klüber Lubrication). The temperature of the aluminum block was maintained at 25.0°C by circulating liquid from a chiller (NCB-1210, Tokyo Rikakikai Co. Ltd.), with temperature readings obtained via Pt resistivity measurements directly on the block.

The polystyrene beads in the water–ethanol mixtures were observed using an inverted optical microscope (Eclipse Ts2, Nikon Solutions Co. Ltd.) equipped with a long-working-distance (LWD) objective lens (40×/0.55, WD 2.1). During the falling-ball experiment, the aluminum sample holder was positioned on the microscope stage, and a series of still images were captured at 50 μm height intervals from the bottom to the top of the holder. Images were acquired every 30 min using a digital camera (DC-9, Panasonic). Sample heights were corrected using the refractive index of the water–ethanol mixtures. The particles were identified and counted using the HoughCircle algorithm from the open-source OpenCV library. Each sample typically contained 400–500 particles. The viscosity of the solution was determined by fitting the observed particle density to the Fokker–Planck equation.

The viscosities of the water–ethanol mixtures were also derived from the Brownian motion of the particles [20]. For the falling-ball experiment, particle movements were recorded with a digital camera positioned 200 μm above the lower coverslip. Three to six movie files, each lasting 3 min, were captured at a frame rate of 29.97 frames per second with a resolution of 1920 × 1080 pixels. The particle positions were analyzed using a modified Python TrackPy library, and the displacements of the particles were calculated.

**III. RESULTS AND DISCUSSION**

The Brownian motion of particles was observed using the same experimental setup as the falling-ball experiment. The 30-min intervals between falling-ball measurements provided sufficient time to record 10-min Brownian motion videos. Notably, the particle size and concentration were consistent for both gravitational and Brownian viscosity measurements.

Fig. 1 depicts the one-dimensional displacement of 1.0 μm particles in pure water at 25.0°C, with interval times corresponding to 10, 20, 30, 60, and 90 frames. Approximately $2.5 \times 10^7$ events were categorized in 0.2 μm displacement increments to construct histograms. The displacement distributions align well with the overlaid Gaussian functions, as indicated by the blue dashed lines representing the fitted Gaussian curves. Since Brownian motion is inherently stochastic, the presence of a normal distribution in the displacement is expected. Logarithmic-scale graphs on the right side of Fig. 1 further validate the agreement of the distribution tail with the Gaussian model. Although a few events ($<10^2$) exhibited statistical noise, the overall agreement with the Gaussian distribution is excellent, spanning over three orders of magnitude in event counts.

Fig. 2 illustrates a displacement histogram for a similar experiment conducted with a 20 wt% ethanol–water solution, under identical conditions to those in Fig. 1. While anomalous Brownian motion has previously been reported in water–ethanol mixtures [20], the displacement data from our experiments fit well with the Gaussian distribution curves. Compared to Fig. 1, the ethanol–water mixture exhibited slightly more frequent statistical noise, likely due to smaller displacements caused by the higher viscosity of the solution. This increase in viscosity reduced the MSD. Additionally, positional errors were more pronounced for smaller displacements, amplifying statistical inaccuracies [41].

These results confirm that Brownian motion displacements, even in ethanol–water mixtures, follow a Gaussian distribution without any anomalies. Consequently, the abnormality in Brownian motion previously observed in such mixtures cannot be attributed to nonlinear processes, such as Lévy flights [42].

For longer interval times, the peak widths of the displacement distributions for both pure and mixed samples increased, accompanied by a rise in variance across the frame intervals. The variance in displacement corresponds to the MSD, which is expressed in Eq. (1) as a function of the diffusion coefficient, itself dependent on the liquid viscosity and particle radius. Fig. 3 demonstrates that the experimental data align well with the Stokes–Einstein equation.

The variance plot for pure water in Fig. 3(a) confirms the agreement between the observed Brownian motion and the straight line calculated using the literature viscosity value for pure water and a particle diameter of 1.035 μm. However, a discrepancy emerges in Fig. 3(b) when comparing the variance derived from the Brownian motion of the aqueous ethanol solution (20 wt% ethanol) with the calculated line based on the literature viscosity. Despite the Gaussian shape of the displacement distribution for the ethanol solution, the variance showed irregularities.

Typically, discrepancies in the Stokes–Einstein equation are addressed by adjusting the effective viscosity or the hydrodynamic radius. To entirely eliminate the observed discrepancy by altering the hydrodynamic radius, the particle diameter would need to be reduced to 0.990 μm from 1.035 μm. However, explaining such a reduction in particle size through molecular solvation effects is implausible. Therefore, the effective viscosity must be reconsidered and is determined to be 1.74 mPa·s, deviating from the literature value of 1.83 mPa·s.

Viscosity measurements of water–ethanol mixtures are traditionally performed using glass capillary viscometers, such as the Ubbelohde viscometer, which measures the time required for a specific liquid volume to flow through a capillary [22]. The falling-ball viscometer is another well-established technique, in which a sphere with a diameter of 1 mm or larger is released into the liquid, and viscosity is calculated based on its descent time. In this study, the falling-ball principle was adapted for the measurement of 1 μm colloidal particles. The motion of the colloid is governed by thermal disturbances, viscosity, and gravitational forces, including buoyancy. This behavior can be modeled using the Fokker–Planck equation [39]:

$$\frac{\partial P(z,t)}{\partial t} = \frac{\partial}{\partial z}\left(\frac{m^*g}{\gamma}P(z,t) + \frac{k_BT}{\gamma}\frac{\partial P(z,t)}{\partial z}\right), \qquad (2)$$

where $P(z,t)$ represents the probability as a function of height $z$ and time $t$, $\gamma\ (=6\pi\eta a)$ represents the Stokes drag experienced by the particle. The term $m^*g$ represents the gravitational force minus the buoyant force, accounting for the effective mass $m^*$. The first term on the right-hand side of Eq. (2) describes one-dimensional gravitational motion, while the second term captures the Brownian motion in all directions. Both terms incorporate the Stokes drag; however, a detailed analysis of Eq. (2) enables the separation of their individual contributions.

Fig. 4 presents the results of the falling-ball experiment for the aqueous solution containing 20 wt% ethanol. Parameters such as viscosity $\eta$, sample height $H$, and a normalization factor required to align the number of particles with the probability defined in Eq. (2) were optimized. The experimental data and the corresponding optimized curves are displayed for each falling period. The optimized parameters successfully describe the data obtained from the falling-ball experiments. The optimized viscosity, 1.82 mPa·s, closely matches the literature value of 1.83 mPa·s, in contrast to the effective viscosity determined from Brownian motion, which is 1.74 mPa·s. Table 1 summarizes the parameters for other water–ethanol mixtures (0, 10, 30, and 40 wt% ethanol). Minor variations in sample height were observed due to the silicon grease sealing; however, these differences remained within acceptable bounds.

At higher ethanol concentrations (>50 wt%), determining the viscosity using the falling-ball method becomes increasingly challenging. This difficulty arises due to the transition from a water–ethanol solution to an ethanol-dominant mixture [32]. The reduced polarizability of the solvent likely contributes to the instability of surface-charged colloids, prolonging the time required for the falling-ball experiment. Efforts are ongoing to measure falling-ball viscosities at higher ethanol concentrations.

In Fig. 5(a), the viscosities obtained from the falling-ball experiments (circles), Brownian motion measurements (crosses), and literature values (solid line) are compared [43,44]. The viscosities for pure water (0 wt% ethanol) determined from both the falling-ball and Brownian motion experiments align well with the literature value, suggesting that the parameters used—particle radius, solvent density (water), colloidal particle density, solvent refractive index, gravity, and temperature—were accurate. These parameters are detailed in Table 1 and its footnote. As the density and refractive index of the aqueous ethanol solution depend on ethanol concentration, their values were derived from literature sources with uncertainties below 0.1%. Consequently, the accuracy of the viscosity measurements for both techniques, approximately 1%, is primarily determined by the experimental methods. The viscosities from the falling-ball experiments matched the literature values within the accuracy limits across all ethanol concentrations. In contrast, the viscosities derived from Brownian motion measurements were consistently lower than the literature values by a statistically significant margin, consistent with prior studies. The observed discrepancies, ranging from 3–5%, align with previously reported values [20,21]. This agreement confirms the reliability of the falling-ball method for viscosity measurement.

In Fig. 5(b), the probability density of particles for the 20 wt% ethanol solution at 420 min of sedimentation is shown (circles), along with a fitted curve (solid line) and a simulation result using the Brownian viscosity (dashed line). Although the difference between the Brownian viscosity and the values obtained from the falling-ball experiment or literature is relatively small (approximately 5%), a noticeable gap exists between the two curves. This gap is better understood by analyzing the falling-ball experiment data. Importantly, Eq. (2) in the Fokker–Planck equation contains two viscosity terms: one associated with the gravitational force and the other with Brownian broadening. While it is theoretically possible to compute two distinct viscosity values, the dominance of the gravitational term allows the data from the falling-ball experiment to be fitted with a single viscosity value.

To address this discrepancy, we turn to the Stokes–Einstein relation [Eq. (1)], which describes the MSD due to Brownian motion. The difference between the predicted and experimental values suggests the presence of an error term in the equation. However, the falling-ball experiment allows for the calculation of viscosity and the complete Stokes drag term, including particle size. Additionally, the displacement distribution from Brownian motion remains entirely Gaussian, with no evidence of nonlinear contributions. These findings suggest that the observed discrepancy is solely attributable to the thermal fluctuation term, $k_B T$.

**IV. MEMORY EFFECT**

Previous studies have examined the coupling between translational and rotational motion of colloidal particles to better understand the unusual Brownian behavior observed in water–ethanol mixtures [20,21]. In this work, the governing equations were refined to satisfy the energy equipartition principle and facilitate more accurate analyses. The detailed mathematical formulation for the coupling of translational and rotational motions is provided in the Supplementary Information [20,21]. Here, we focus on the Brownian motion of a spherical particle with radius $a$, as depicted in Fig. 6. Given the significant variation in viscosity within water–ethanol mixtures depending on ethanol concentration, it is reasonable to expect that concentration fluctuations could manifest as viscosity fluctuations. These viscosity differences can induce translational forces through rotational friction [20,45]:

$$F'_x = 3\pi a^2 \Delta\eta_y \omega_z, \qquad (3)$$

where $\omega_z$ represents the angular velocity along the $z$-axis. A similar translational force arises due to the relationship between $\Delta\eta_z$ and $\omega_y$. These forces are incorporated into the one-dimensional Langevin equation for motion in the $x$-direction, which accounts for the mass $m$ of the particle, thermal fluctuations $R(t)$, and the conversion forces:

$$m\frac{dx^2}{dt^2} = -\gamma\frac{dx}{dt} + 3\pi a^2 \Delta\eta_y \omega_z - 3\pi a^2 \Delta\eta_z \omega_y + R(t). \qquad (4)$$

Typically, the Langevin equation does not include the conversion force $F'_x$. Notably, these conversion forces interact with thermal fluctuations, introducing a memory effect. In this context, thermal fluctuations are categorized into two types: those associated with rotational Brownian motion memory and pure random impacts without memory. The pure random thermal fluctuation, $R(t)$, satisfies the following relations:

$$\langle R(t) \rangle = 0$$
$$\langle R(t)R(t') \rangle = 2B\delta(t - t'). \tag{5}$$

Here $\langle \cdot \rangle$ represents statistical averaging. Using the energy equipartition principle, the amplitude of random fluctuations can be expressed as:

$$B = \left\{ 1 - \frac{15}{52} \frac{\langle \Delta \eta(t)^2 \rangle}{\eta^2} \right\} \gamma k_B T, \tag{6}$$

where $\langle \Delta \eta(t)^2 \rangle = \langle \Delta \eta_y(t)^2 \rangle = \langle \Delta \eta_z(t)^2 \rangle$. Rotational motion of the particles undergoes Brownian fluctuations, with the time correlation function of $\omega \, (= \omega_y \text{ or } \omega_z)$, which can be expressed as follows:

$$\langle \omega(t)\omega(t') \rangle = \frac{k_B T}{I} e^{-\frac{\gamma'}{I}|t-t'|}, \tag{7}$$

where $I$ and $\gamma' (= 8\pi a^3 \eta)$ denote the moment of inertia of the particle and rotational viscous coefficient, respectively. Solving the Langevin equation with Eqs. (6) and (7) yields the following expression for the MSD:

$$\langle \Delta x^2 \rangle = \frac{2k_B T}{\gamma} \left\{ 1 + \frac{9}{104} \frac{\langle \Delta \eta(t)^2 \rangle}{\eta^2} \right\} \cdot t. \tag{8}$$

The $\langle \Delta x^2 \rangle$ increases slightly due to the second term in parentheses, reflecting the memory effect from the conversion of rotational to translational motion. In the generalized Langevin framework, non-Markovian processes with memory effects can be approximated as Markovian processes with an effective mass [46]. This formulation predicts an increased amplitude of thermal fluctuations compared to normal thermal noise, while preserving the Gaussian displacement distribution typically observed.

## V. CONCLUSION

Viscous forces acting on colloids in water-ethanol mixtures, as measured by falling-ball experiments, align with the values for bulk liquids. However, Brownian motion experiments revealed significantly larger translational displacements, suggesting a nonlinear process involving memory effects. The thermal fluctuation is not merely a memoryless random collision but also includes a conversion from rotational Brownian motion under inhomogeneous viscosity. The relaxation time of rotational Brownian motion ($\tau_r = I/\gamma'$) is typically approximately 0.02 μs. This brief, non-zero duration can accumulate over a 1-s period, resulting in notable translational displacement. Although the long-term behavior may resemble a linear process, the physical quantities involved must account for the nonlinear nature of the memory effect.

## ACKNOWLEDGMENT

This study was supported by a Grant-in-Aid for Scientific Research (C) from the Japan Society for the Promotion of Science (JSPS), KAKENHI (Grant No. JP17K05765).

Table 1. Optimized parameters obtained from the falling-ball measurements. The refractive index and density of the solvents are based on previous studies [43].

| ethanol concentration [wt%] | 0 | 10 | 20 | 30 | 40 |
|---|---|---|---|---|---|
| viscosity (gravitational) [mPa·s] | 0.898 | 1.33 | 1.82 | 2.20 | 2.39 |
| height [mm] | 1.11 | 1.14 | 1.11 | 1.12 | 1.16 |
| refractive index | 1.33252 | 1.33899 | 1.34600 | 1.35213 | 1.35647 |
| density [g/cm$^3$] | 0.9969 | 0.9807 | 0.9663 | 0.9504 | 0.9316 |
| viscosity (Brownian) [mPa·s] | 0.894 | 1.28 | 1.74 | 2.14 | 2.31 |

Constant values for the following parameters are utilized to calculate all concentrations:
- Particle diameter: 1.035 μm
- Polystyrene density: 1.043 g/cm$^3$
- Absolute temperature: 298.15 K
- Acceleration due to gravity: 9.798 m/s$^2$

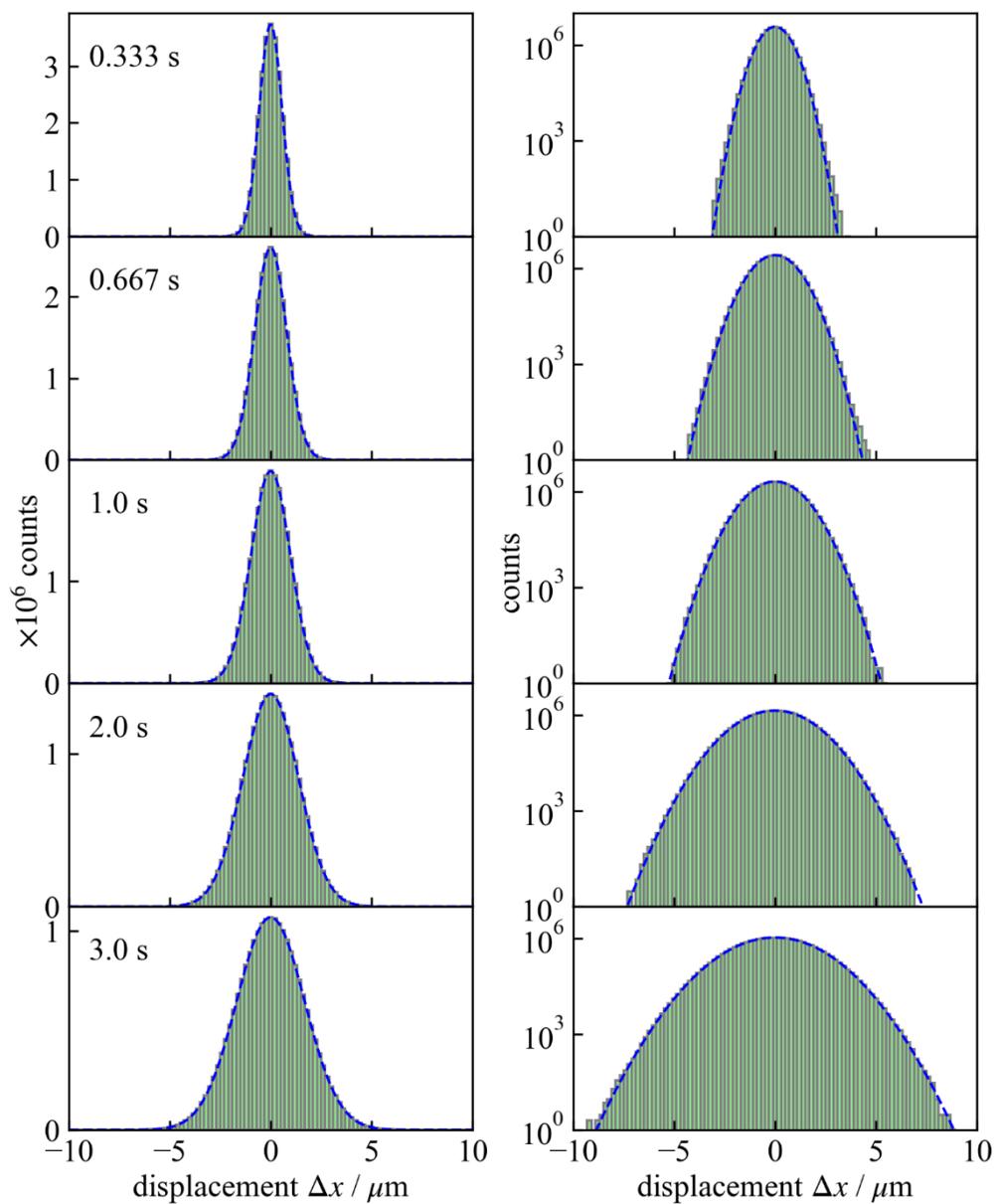

FIG. 1. Brownian motion displacements of 1.0 μm beads in pure water at 25.0°C. Gaussian functions provide an excellent fit to the histograms for each lag time (left: linear scale), with the tails of the histograms also fitting well to the Gaussian distributions (right: logarithmic scale).

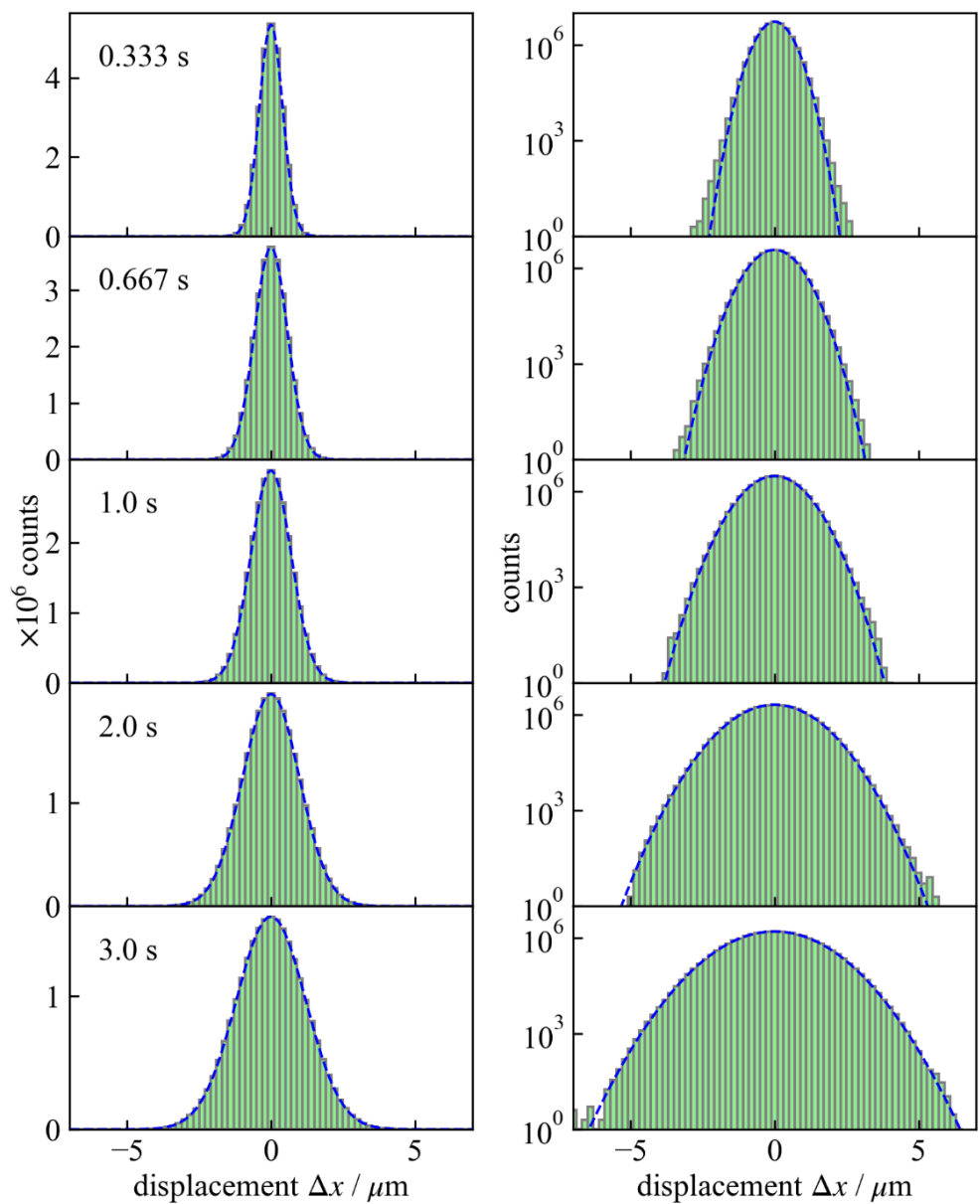

FIG. 2. Brownian motion displacements of 1.0 μm beads in an aqueous solution containing 20 wt% ethanol at 25.0°C. A Gaussian function provides an excellent fit to the histogram for each lag time (left: linear scale), with the tails of the histograms also fitting well to the Gaussian distributions (right: log scale).

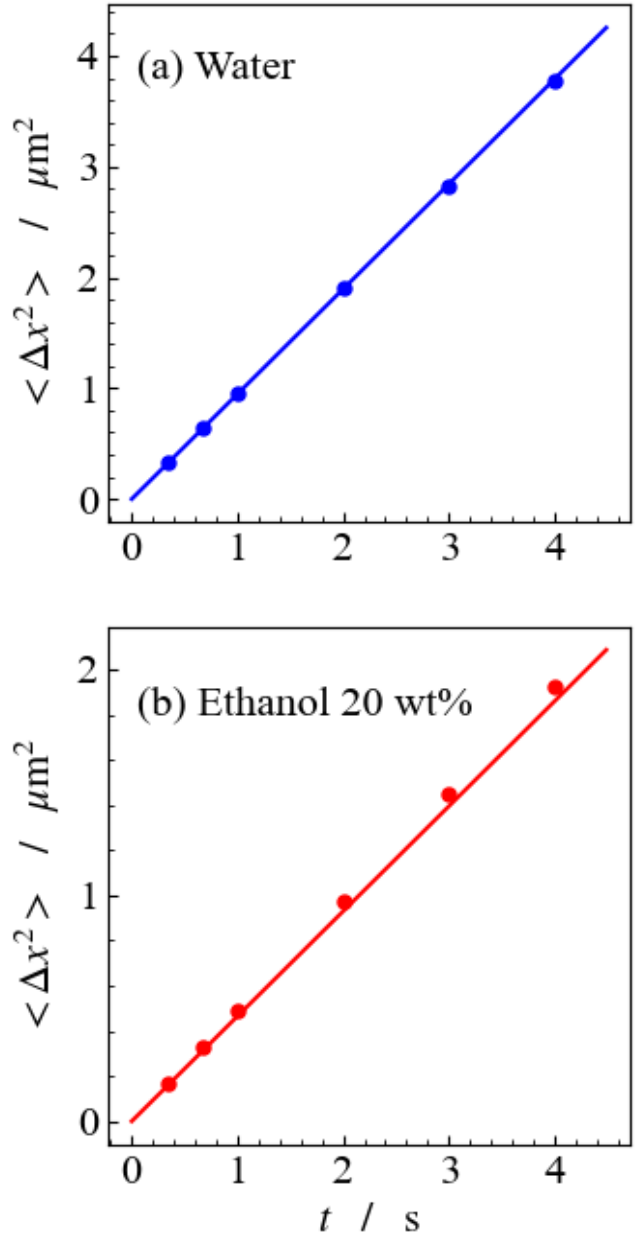

FIG. 3. Variance of Brownian motion displacement, i.e., MSD. The MSD for pure water (a) shows good agreement with the Stokes–Einstein relation. However, the calculated MSD for the 20 wt% ethanol aqueous solution (b) deviates significantly from the value obtained using the bulk liquid viscosity reported in the literature.

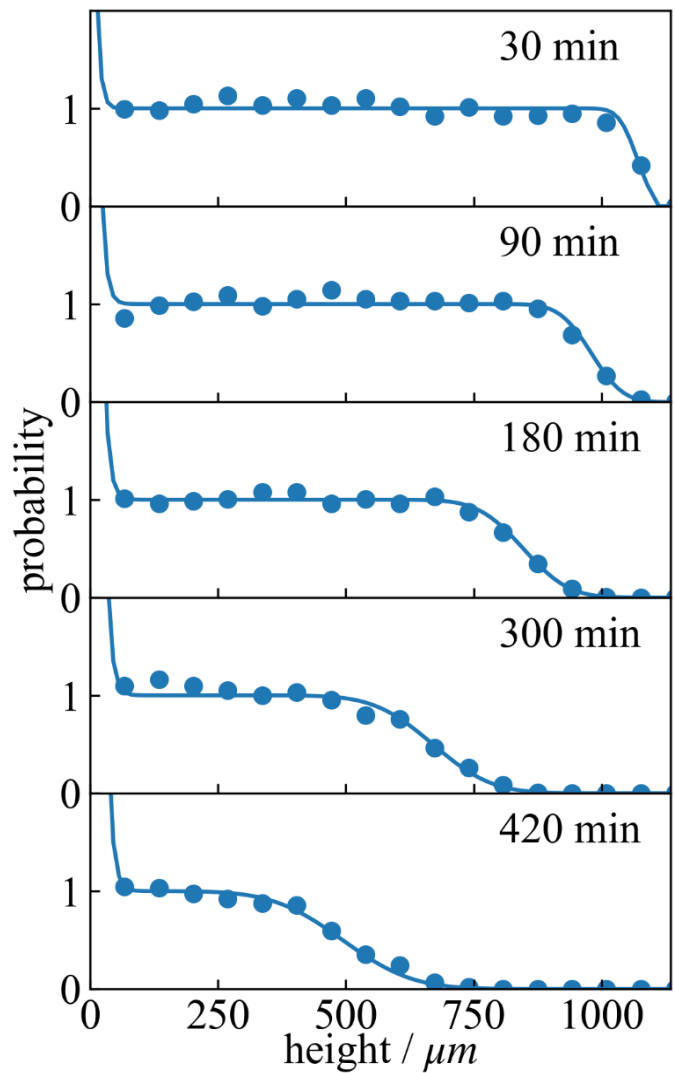

FIG. 4. Results of the falling-ball experiment for viscosity determination. Normalized particle counts are plotted for the aqueous solution containing 20 wt% ethanol at 25.0°C. The solid lines represent the theoretical probabilities generated using optimized parameters, including the optimized viscosity value.

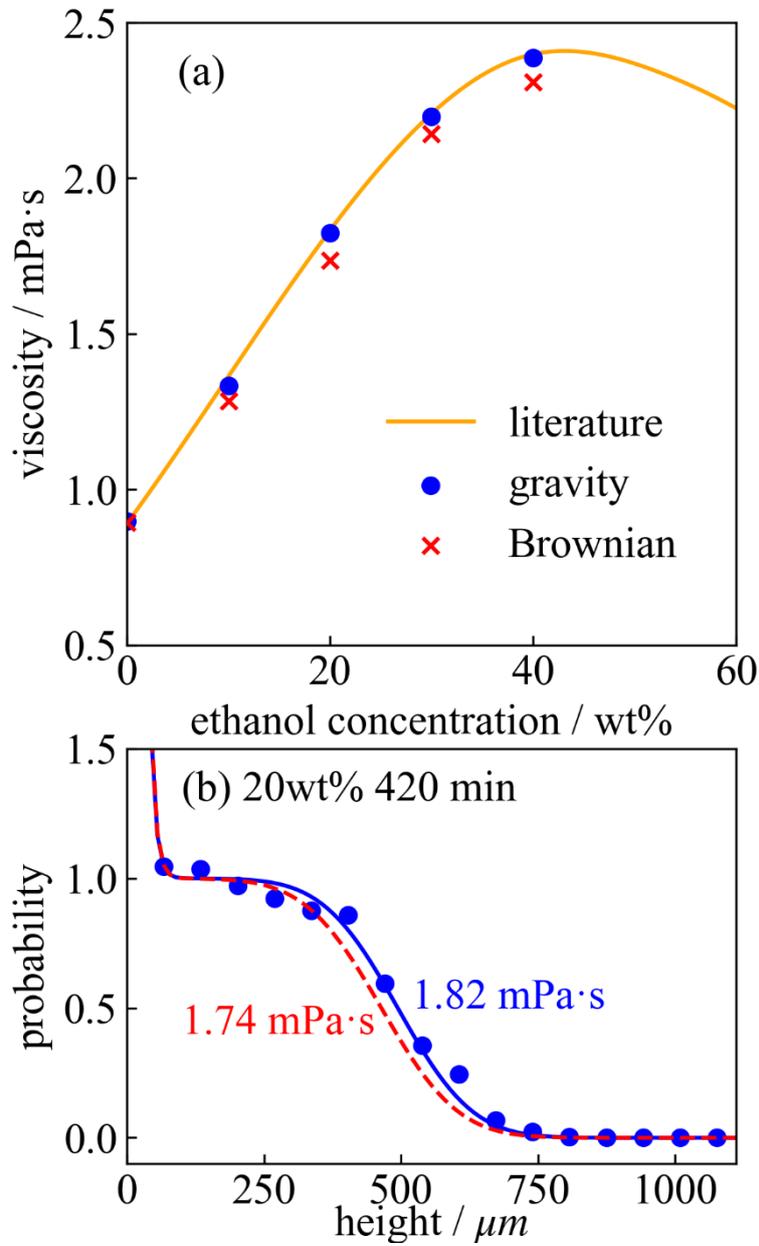

FIG. 5. Comparisons of viscosities determined from the falling-ball experiment, Brownian motion measurements, and previously reported studies. (a) Viscosity dependence on ethanol concentration. The viscosities (circles) from the falling-ball experiment align well with the literature values. The viscosities calculated from the Brownian motion measurements (crosses) are slightly lower than the literature values, except for the result obtained from pure water. (b) Results from the falling-ball experiment in a 20 wt% ethanol solution. The plot shows the observed number of particles at 420 min at each height, with the solid blue line representing a fit to the data based on optimized viscosity values. In contrast, the dashed red line represents simulated data using the Brownian viscosity. The calculated functions exhibit a noticeable difference.

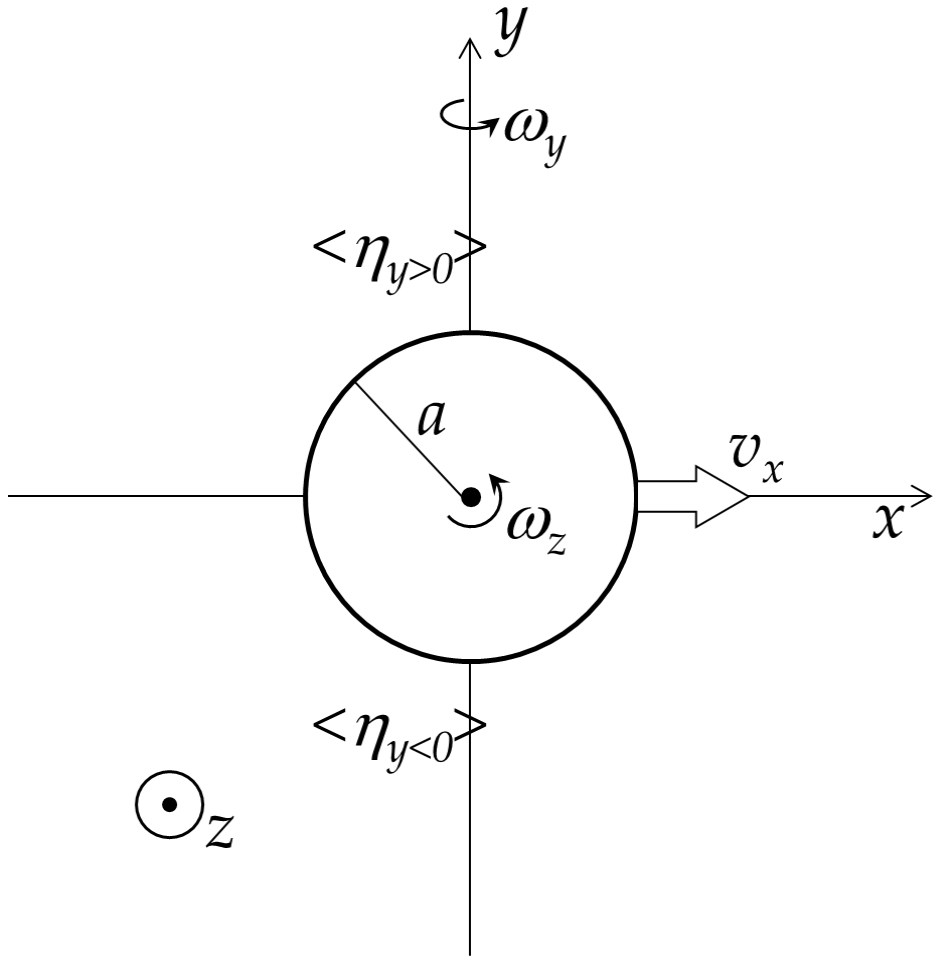

FIG. 6. Sphere in viscosity fluctuation. The difference in viscosity fluctuation between the *xy*-plane ($\Delta\eta_z$) and *xz*-plane ($\Delta\eta_y$) induces movement along the x-axis due to the rotation around the y-axis ($\omega_y$) and z-axis ($\omega_z$) respectively.

Supplementary Information

Memory effect by coupling between translational and rotational
Brownian motion in water–ethanol mixtures

Ken Judai[1]*, Satoshi Shibuta[1,2], Kazuki Furukawa[1,3]

[1]Department of Physics, College of Humanities and Sciences, Nihon University, Sakurajosui 3-25-40, Setagaya-ku, Tokyo 156-8550, Japan

[2]National Institute of Technology, Kagoshima College, Shinko Hayato 1460-1, Kirishima, Kagoshima 899-5193, Japan

[3]National Institute of Technology, Ariake College, Higashihagio-Machi 150, Omuta, Fukuoka 836-8585, Japan

**Additional experimental results**

Below, we provide a detailed description of the experimental results for the sedimentation process. Fig. S1 illustrates the schematic experimental setup. Polystyrene beads, each 1.0 μm in diameter, were suspended in water within a homemade sample holder, which was then placed on the stage of an inverted optical microscope. Photographs were taken while varying the height of the sample over time. Fig. S2 presents cropped sections (2000 × 1000 pixels) from the original 5184 × 3888-pixel images. These photographs were captured at different heights from the bottom of the sample: (a) 66.626 μm, (b) 399.756 μm, (c) 799.512 μm, and (d) 1066.016 μm. Blue circles in the images indicate particles detected automatically using the HoughCircle algorithm from OpenCV. Photograph (b), taken at an intermediate height, shows the typical initial particle concentration. After sedimentation, the number of particles at lower positions increased, as shown in Fig. S2(a), while fewer particles were observed at higher positions, as evident in Fig. S2(c). Notably, almost no particles were detected at the highest position, as indicated in Fig. S2(d).

Fig. S3 shows the particle count as a function of height. The experimental height was determined by the position of the objective lens, which moved in increments of 50 μm in air. However, due to the refractive index of water (1.33252), the focus position shifts by 66.626 μm (1.33252 times the air distance). Refractive index corrections were also applied for certain concentrations of the ethanol–water mixture. Typically, the initial particle count ranged from 400 to 500. The Fokker–Planck equation was used to calculate the probability density of the particles. In Eq. (S13), $P(z,t)$ represents the probability density integrated from 0 to $H$ (the sample height) and normalized to 1. To fit the experimental data to the Fokker–Planck equation, a scaling factor was introduced. This factor accounts for both the number of particles (400–500) and the conversion from density probability, which is scaled by the inverse of the height $1/H$.

Fig. S4 shows the fitting curves obtained using the Fokker-Planck equation for all sedimentation times. The data were normalized based on the number of particles, setting the initial probability to 1 for all positions, rather than normalizing by the probability density ($1/H$). The simulated curves closely match all the experimental data points. Although small discrepancies between the predicted and experimental results could arise due to factors such as vibrations of the optical table or solvent flow during the falling-ball experiment, it is important to note that we obtained reliable viscosity values under carefully controlled experimental conditions.

**Fokker−Planck equation**

Particles within the water and ethanol mixtures undergo motion due to gravitational sedimentation and random thermal forces, as described by the Langevin equation [Eq. (4)]. The movement of a particle along the gravitational direction can be conveniently described in terms of the probability density $P(z,t)$, which represents the probability of finding a particle within the interval $z$ to $z + dz$ at time $t$. The evolution of this probability function is governed by the Fokker–Planck equation, which can be expressed as follows:

$$\frac{\partial P(z,t)}{\partial t} = \frac{\partial}{\partial z}\left(\frac{m^*g}{\gamma}P(z,t) + D\frac{\partial P(z,t)}{\partial z}\right) = -\frac{\partial J(z,t)}{\partial z}, \quad (S1)$$

where $\gamma$ ($= 6\pi\eta a$) denotes the Stokes' drag, $m^*g$ denotes the gravitational force including the buoyant force, $D$ ($= k_BT/\gamma$) denotes the diffusion coefficient, and $J(z,t)$ denotes the probability current [39]. We assume that there exists a solution to Eq. (S1) in the form of $P_\lambda(z,t) = e^{-\lambda t}F_\lambda(z)$ ($\lambda \geq 0$). This solution satisfies the following equation:

$$-\lambda F_\lambda(z) = \frac{m^*g}{\gamma}\frac{dF_\lambda(z)}{dz} + D\frac{d^2F_\lambda(z)}{dz^2}. \quad (S2)$$

The solution to this non-homogeneous differential equation [Eq. (S2)] is as follows:

$$F_\lambda(z) = e^{-\alpha z}\{A_\lambda \cos(\alpha R_\lambda z) + B_\lambda \sin(\alpha R_\lambda z)\}, \quad (S3)$$

where $\alpha = \frac{m^*g}{2D\gamma}$ and $R_\lambda = \pm\sqrt{\frac{4D\gamma^2\lambda}{m^{*2}g^2} - 1}$. The particle probability current $J$ is given by:

$$J_\lambda(z,t) = \frac{m^*g}{\gamma}P_\lambda(z,t) + D\frac{\partial P_\lambda(z,t)}{\partial z}$$
$$= -De^{-\lambda t}e^{-\alpha z}\alpha\{(A_\lambda + R_\lambda B_\lambda)\cos(\alpha R_\lambda z) + (B_\lambda - R_\lambda A_\lambda)\sin(\alpha R_\lambda z)\}. \quad (S4)$$

Next, we determine $\lambda, A_\lambda$ and ,$B_\lambda$ under the experimental conditions. The water–ethanol solvent mixture was sealed with coverslips, and particles could not pass through the top ($z = H$) or bottom ($z = 0$) coverslips. This results in no particle current at these points $J_\lambda(H,t) = 0, J_\lambda(0,t) = 0$. From the boundary condition $J_\lambda(0,t) = 0$, we obtain the relation $A_\lambda = -R_\lambda B_\lambda$. Therefore, Eq. (S4) simplifies to:

$$J_\lambda(z,t) = -De^{-\lambda t}e^{-\alpha z}\alpha B_\lambda(1 + R_\lambda^2)\sin(\alpha R_\lambda z). \tag{S5}$$

Regarding the boundary condition $J_\lambda(H,t) = 0$, we examine two distinct scenarios: the stationary state $\lambda = 0$ and the non-stationary state $\lambda \neq 0$.

When $\lambda = 0$, i.e., in the stationary state, the probability density becomes independent of time. To satisfy the condition $J_\lambda(H,t) = 0$, the coefficient in Eq. (S5), denoted as $(1 + R_\lambda^2)$, must be equal to zero, which gives $R_0 = \pm i$. Consequently, we determine $F_0(z) = iB_0\exp[-2\alpha z] = iB_0\exp[-m^*gz/k_BT]$. When we normalize the solution over the range $0 \leq z \leq H$, we obtain:

$$P_0(z) = F_0(z) = \frac{m^*g}{k_BT}\frac{e^{-m^*gz/k_BT}}{1 - e^{-m^*gH/k_BT}}. \tag{S6}$$

For $\lambda \neq 0$, $J_\lambda(H,t) = 0$ requires that $\sin(\alpha R_\lambda z)$ in Eq. (S5) equals zero, i.e., $\alpha R_\lambda H = n\pi$ ($n = 1, 2, \cdots \infty$). Therefore, we obtain:

$$\lambda_n = \frac{m^{*2}g^2}{4D\gamma^2} + \frac{n^2\pi^2 D}{H^2}, \tag{S7}$$

Although the boundary condition discretizes $\lambda$ into integer values, it is essential to note that the solution requires an infinite series of terms $\lambda_n$. The solution to the Fokker–Planck equation can be expressed as:

$$P(z,t) = P_0(z) + \sum_{n=1}^{\infty} e^{-\lambda_n t}F_{\lambda_n}(z), \tag{S8}$$

or, using Eq. (S2) and (S7),

$$P(z,t) = P_0(z) + \sum_{n=1}^{\infty} A_n e^{-\left(\frac{m^{*2}g^2}{4\gamma k_B T} + \frac{k_B T n^2 \pi^2}{\gamma H^2}\right)t} e^{-\frac{m^*gz}{2k_BT}}\left\{\cos\left(\frac{n\pi z}{H}\right) - \frac{m^*gH}{2k_BTn\pi}\sin\left(\frac{n\pi z}{H}\right)\right\}. \tag{S9}$$

To derive the coefficients $A_n$, we define $\mathcal{H} := \frac{m^*g}{\gamma}\frac{\partial}{\partial z} + D\frac{\partial^2}{\partial z^2}$. $\mathcal{H}$ is a Hamiltonian operator satisfying $\mathcal{H}\Phi_n(z) = -\lambda_n\Phi_n(z)$ with eigenvalues $\lambda_n$ and a complete set of eigenfunctions $\Phi_n(z)$. The probability $P(z,t)$ can be expanded in terms of these eigenfunctions as follows:

$$P(z,t) = \Phi_0^2(z) + \sum_{n=1}^{\infty} C_n e^{-\lambda_n t}\Phi_0(z)\Phi_n(z), \tag{S10}$$

where $\Phi_0(z)$ is the stationary state distribution, defined as $\Phi_0(z) = \sqrt{2\alpha/(1 - e^{-2\alpha H})}\,e^{-\alpha z}$, and $\Phi_n(z)$ is expressed in terms of $\alpha = m^*g/2k_BT$ by:

$$\Phi_n(z) = \sqrt{\frac{2}{H}}\frac{n\pi}{\sqrt{\alpha^2 H^2 + n^2\pi^2}}\left\{\cos\left(\frac{n\pi z}{H}\right) - \frac{\alpha H}{n\pi}\sin\left(\frac{n\pi z}{H}\right)\right\}. \tag{S11}$$

Since the eigenfunctions $\Phi_n(z)$ form a complete orthonormal basis, $\int_0^H \Phi_i(z)\Phi_j(z)dz = \delta_{ij}$, we can determine the coefficients using the initial distribution, $P(z,0) = 1/H$:

$$C_n = \int_0^H dz \frac{\Phi_n(z)}{\Phi_0(z)} \frac{1}{H}$$

$$= 2\alpha H n\pi \sqrt{\frac{1-e^{-2\alpha H}}{\alpha H}} \frac{(e^{\alpha H}\cos(n\pi)-1)}{(\alpha^2 H^2 + n^2\pi^2)^{3/2}}. \tag{S12}$$

Finally, we obtain the particle probability density using Eqs. (S10) and (S12) as follows:

$$P(z,t) = \frac{2\alpha}{1-e^{-2\alpha H}} e^{-2\alpha z} + \sum_{n=1}^{\infty} \frac{(e^{\alpha H}\cos(n\pi)-1)4\alpha n^2\pi^2}{(\alpha^2 H^2 + n^2\pi^2)^2} \left\{\cos\left(\frac{n\pi z}{H}\right) - \frac{\alpha H}{n\pi}\sin\left(\frac{n\pi z}{H}\right)\right\} e^{-(\lambda_n t + \alpha z)}. \tag{S13}$$

The density of particles in the falling-ball experiment was nonlinearly optimized using Eq. (S13).

**Coupling of translation and rotation with Memory Effect**

The mean squared displacement (MSD) of Brownian motion in the presence of inhomogeneous viscosity is influenced by the coupling between translational and rotational movements. Here, we focus on the motion along the x-axis (see Fig. 6). The Brownian motion along the x-axis is affected by the inhomogeneous viscosity within the divided *xz*- and *xy*-planes. The surface-averaged viscosity inhomogeneities acting on the Brownian particle are defined by the viscosity difference between regions where y>0 and y<0, or z>0 and z<0, respectively:

$$\Delta\eta_y(t) := \langle\eta_{y>0}\rangle - \langle\eta_{y<0}\rangle$$

$$\Delta\eta_z(t) := \langle\eta_{z>0}\rangle - \langle\eta_{z<0}\rangle.$$

The Navier–Stokes equation is employed to determine the viscosity of a fluid under the Stokes approximation, which assumes incompressibility. Considering spherical coordinates $(r, \theta, \phi)$ with respect to the Cartesian *x*-, *y*-, and *z*-axes, the fluid velocities can be expressed as:

$$v_r = v_\theta = 0,$$

$$v_\phi = \left(\frac{a^3}{r^2}\right)\omega_z \sin\theta,$$

where $\omega_z$ represents the angular velocity of the sphere along the z-axis. Additionally, the frictional force acting on a unit area is given by:

$$\sigma_{r\phi} = \eta\left(\frac{\partial v}{\partial r} - \frac{v}{r}\right)_{r=a} = -3\eta\omega_z \sin\theta.$$

In this context, we investigate the conversion of the viscosity difference in the *xz*-plane ($\Delta\eta_y$) from *z*-axis rotation ($\omega_z$) to *x*-axis translation. This viscosity variation results in a translational force, $F_x'$, as there is no counterbalancing effect in the *xz*-plane.

$$F_x' = \int_0^\pi \int_0^\pi \{-3\langle\eta_{y>0}\rangle\omega_z \sin\theta \, (-\sin\phi) a^2 \sin\theta\} d\phi d\theta$$

$$+ \int_0^\pi \int_\pi^{2\pi} \{-3\langle\eta_{y<0}\rangle\omega_z \sin\theta \, (-\sin\phi) a^2 \sin\theta\} d\phi d\theta$$

$$= 3\pi a^2 \langle \eta_{y>0}\rangle \omega_z - 3\pi a^2 \langle \eta_{y<0}\rangle \omega_z$$
$$= 3\pi a^2 \Delta\eta_y \omega_z.$$

In the case of translation along the *x*-axis, we also observe the conversion of the viscosity difference in the *xy*-plane ($\Delta\eta_z$) resulting from *y*-axis rotation. The additional force, $F_x$, representing the transition from rotation to translation, can be expressed as the sum of two components:

$$F_x = 3\pi a^2 \Delta\eta_y \omega_z - 3\pi a^2 \Delta\eta_z \omega_y. \tag{S14}$$

These additional forces modify the Langevin equation for the particle's position along the x-axis, as follows:

$$m\frac{dv_x(t)}{dt} = -\gamma\, v_x(t) + 3\pi a^2 \Delta\eta_y(t)\omega_z(t) - 3\pi a^2 \Delta\eta_z(t)\omega_y(t) + R(t), \tag{S15}$$

where $\gamma\, (= 6\pi\eta a)$ is the translational Stokes coefficient, and $R(t)$ is the thermal random force. The thermal random force introduces fluctuations in Brownian motion due to solvent molecular collisions. Since there is no memory effect associated with these collisions, it is expressed as:

$$\langle R(t)\rangle = 0$$
$$\langle R(t)R(t')\rangle = 2B\delta(t-t').$$

In contrast to thermal collisions, the converting force from rotation to translation contains a memory effect due to the Brownian rotational relaxation time. Since both $F_x$ and $R(t)$ are thermal fluctuations, the amplitude of the thermal random force, $B$, must be determined from the energy equipartition principle.

Integrating Eq. (S15) yields the velocity of the particle:

$$v_x(t) = v_x(0)e^{-\frac{\gamma}{m}t} + \int_0^t dt'\, e^{-\frac{\gamma}{m}(t-t')} \left\{ \frac{3\pi a^2 \Delta\eta_y(t')\omega_z(t')}{m} - \frac{3\pi a^2 \Delta\eta_z(t')\omega_y(t')}{m} + \frac{R(t')}{m} \right\}$$

where $v(0)$ denotes an initial velocity. Since the observed time is much longer than the viscous force relaxation time ($m/\gamma$), the first term can be neglected as follows:

$$v_x(t) \cong \int_0^t dt'\, e^{-\frac{\gamma}{m}(t-t')} \left\{ \frac{3\pi a^2 \Delta\eta_y(t')\omega_z(t')}{m} - \frac{3\pi a^2 \Delta\eta_z(t')\omega_y(t')}{m} + \frac{R(t')}{m} \right\}. \tag{S16}$$

Consequently, the ensemble average of squared velocity can be calculated as:

$$\langle v_x(t)^2\rangle = \int_0^t dt' \int_0^t dt''\, e^{-\frac{\gamma}{m}(t-t')} e^{-\frac{\gamma}{m}(t-t'')} \left\{ \frac{(3\pi a^2)^2}{m^2} \langle \Delta\eta_y(t')\Delta\eta_y(t'')\rangle \langle \Delta\omega_z(t')\Delta\omega_z(t'')\rangle \right.$$
$$\left. + \frac{(3\pi a^2)^2}{m^2} \langle \Delta\eta_z(t')\Delta\eta_z(t'')\rangle \langle \Delta\omega_y(t')\Delta\omega_y(t'')\rangle + \frac{\langle R(t')R(t'')\rangle}{m^2} \right\}.$$

Assuming that the viscosity fluctuations occur on a timescale much slower than the rotational Brownian motion, we treat it as $\langle \Delta\eta_y(t')\Delta\eta_y(t'')\rangle = \langle \Delta\eta_y(t)^2\rangle$. The integral was then performed for each term:

$$\int_0^t dt' \int_0^t dt''\, e^{-\frac{\gamma}{m}(t-t')} e^{-\frac{\gamma}{m}(t-t'')} \frac{(3\pi a^2)^2}{m^2} \langle \Delta\eta_y(t')\Delta\eta_y(t'')\rangle \langle \Delta\omega_z(t')\Delta\omega_z(t'')\rangle$$

$$= \frac{(3\pi a^2)^2}{m^2} \langle \Delta \eta_y(t)^2 \rangle e^{-\frac{2\gamma}{m}t} \int_0^t dt' \int_0^t dt'' \, e^{\frac{\gamma}{m}t'} e^{\frac{\gamma}{m}t''} \frac{k_B T}{I} e^{-\frac{\gamma'}{I}|t'-t''|}$$

$$= \frac{(3\pi a^2)^2}{m^2} \frac{k_B T}{I} \langle \Delta \eta_y(t)^2 \rangle e^{-\frac{2\gamma}{m}t} \int_0^t dt' \left\{ \int_0^{t'} dt'' \, e^{\frac{\gamma}{m}t'} e^{\frac{\gamma}{m}t''} e^{-\frac{\gamma'}{I}(t'-t'')} \right.$$

$$\left. + \int_{t'}^t dt'' \, e^{\frac{\gamma}{m}t'} e^{\frac{\gamma}{m}t''} e^{-\frac{\gamma'}{I}(t''-t')} \right\}$$

$$= \frac{(3\pi a^2)^2}{m^2} \frac{k_B T}{I} \langle \Delta \eta_y(t)^2 \rangle \frac{1}{\left(\frac{\gamma}{m}\right)^2 - \left(\frac{\gamma'}{I}\right)^2} \left\{ 1 - 2e^{-\left(\frac{\gamma}{m}+\frac{\gamma'}{I}\right)t} - e^{-\frac{2\gamma}{m}t} - \frac{m\gamma'}{I\gamma}\left(1 - e^{-\frac{2\gamma}{m}t}\right) \right\},$$

where $\gamma' (= 8\pi\eta a^3)$ is the rotational Stokes coefficient, and $I (= \frac{2}{5} m a^2)$ is the moment of inertia of the particle. The observation time for Brownian motion, typically exceeding 0.1 s, is much longer than the Brownian relaxation times for translation (0.06 μs) and rotation (0.02 μs). Consequently, the exponential terms can be neglected as follows:

$$\cong \frac{(3\pi a^2)^2}{m^2} \frac{k_B T}{I} \langle \Delta \eta_y(t)^2 \rangle \frac{1}{\left(\frac{\gamma}{m}\right)^2 - \left(\frac{\gamma'}{I}\right)^2} \left\{ 1 - \frac{m\gamma'}{I\gamma} \right\}.$$

The relation of $\left(\frac{\gamma'}{I}\right) = \frac{10}{3}\left(\frac{\gamma}{m}\right)$ makes the equation simpler as:

$$= \frac{15}{104} \frac{1}{m} \frac{\langle \Delta \eta_y(t)^2 \rangle}{\eta^2} k_B T. \tag{S17}$$

Another term was also evaluated:

$$\int_0^t dt' \int_0^t dt'' \, e^{-\frac{\gamma}{m}(t-t')} e^{-\frac{\gamma}{m}(t-t'')} \frac{\langle R(t') R(t'') \rangle}{m^2} = \int_0^t dt' \int_0^t dt'' \, e^{-\frac{\gamma}{m}(t-t')} e^{-\frac{\gamma}{m}(t-t'')} \frac{2B\delta(t'-t'')}{m^2}$$

$$= \frac{B}{m\gamma}\left(1 - e^{-\frac{2\gamma}{m}t}\right) \cong \frac{B}{m\gamma}. \tag{S18}$$

Finally, the Eqs. (S17) and (S18) are combined as follows:

$$\langle v_x(t)^2 \rangle = \frac{15}{104} \frac{1}{m} \frac{\langle \Delta \eta_y(t)^2 \rangle}{\eta^2} k_B T + \frac{15}{104} \frac{1}{m} \frac{\langle \Delta \eta_z(t)^2 \rangle}{\eta^2} k_B T + \frac{B}{m\gamma}. \tag{S19}$$

The averaged viscosity inhomogeneity should be $\langle \Delta \eta_y(t)^2 \rangle = \langle \Delta \eta_z(t)^2 \rangle \coloneqq \langle \Delta \eta(t)^2 \rangle$. The equipartition energy principle is also considered.

$$\frac{1}{2} m \langle v_x(t)^2 \rangle = \frac{15}{104} \frac{\langle \Delta \eta(t)^2 \rangle}{\eta^2} k_B T + \frac{B}{2\gamma} = \frac{1}{2} k_B T$$

$$B = \left\{ 1 - \frac{15}{52} \frac{\langle \Delta \eta(t)^2 \rangle}{\eta^2} \right\} \gamma k_B T \tag{S20}$$

The thermal random force, $B$, decreases with the second term in the parentheses to satisfy the energy equipartition principle, as the force resulting from the conversion of Brownian rotation in inhomogeneous viscosity induces additional translational motion.

Next, we will derive the MSD of the translational Brownian motion.

$$v_x(t) = \int_0^t dt' \, e^{-\frac{\gamma}{m}(t-t')} \left\{ \frac{3\pi a^2 \Delta\eta_y(t')\omega_z(t')}{m} - \frac{3\pi a^2 \Delta\eta_z(t')\omega_y(t')}{m} + \frac{R(t')}{m} \right\}$$

The further integral can produce the displacement of the particle as:

$$\Delta x(t) = \int_0^t dt' \int_0^{t'} dt'' \, e^{-\frac{\gamma}{m}(t'-t'')} \left\{ \frac{3\pi a^2 \Delta\eta_y(t'')\omega_z(t'')}{m} - \frac{3\pi a^2 \Delta\eta_z(t'')\omega_y(t'')}{m} + \frac{R(t'')}{m} \right\}.$$

Changing the order of double integrals, we get:

$$\Delta x(t) = \int_0^t dt'' \int_{t''}^t dt' \, e^{-\frac{\gamma}{m}(t'-t'')} \left\{ \frac{3\pi a^2 \Delta\eta_y(t'')\omega_z(t'')}{m} - \frac{3\pi a^2 \Delta\eta_z(t'')\omega_y(t'')}{m} + \frac{R(t'')}{m} \right\}$$

$$= -\frac{m}{\gamma} e^{-\frac{\gamma}{m}t} \int_0^t dt'' \, e^{-\frac{\gamma}{m}t''} \left\{ \frac{3\pi a^2 \Delta\eta_y(t'')\omega_z(t'')}{m} - \frac{3\pi a^2 \Delta\eta_z(t'')\omega_y(t'')}{m} + \frac{R(t'')}{m} \right\}$$

$$+ \frac{m}{\gamma} \int_0^t dt'' \left\{ \frac{3\pi a^2 \Delta\eta_y(t'')\omega_z(t'')}{m} - \frac{3\pi a^2 \Delta\eta_z(t'')\omega_y(t'')}{m} + \frac{R(t'')}{m} \right\}.$$

The exponential terms can be neglected and computed as follows:

$$\Delta x(t) \cong \frac{m}{\gamma} \int_0^t dt'' \left\{ \frac{3\pi a^2 \Delta\eta_y(t'')\omega_z(t'')}{m} - \frac{3\pi a^2 \Delta\eta_z(t'')\omega_y(t'')}{m} + \frac{R(t'')}{m} \right\}. \tag{S21}$$

The MSD can be expressed as follows:

$$\langle \Delta x(t)^2 \rangle = \left(\frac{m}{\gamma}\right)^2 \int_0^t dt' \int_0^t dt'' \left\{ \frac{(3\pi a^2)^2}{m^2} \langle \Delta\eta_y(t')\Delta\eta_y(t'')\rangle \langle \Delta\omega_z(t')\Delta\omega_z(t'')\rangle \right.$$

$$\left. + \frac{(3\pi a^2)^2}{m^2} \langle \Delta\eta_z(t')\Delta\eta_z(t'')\rangle \langle \Delta\omega_y(t')\Delta\omega_y(t'')\rangle + \frac{\langle R(t')R(t'')\rangle}{m^2} \right\}$$

The integral was performed for each term as follows:

$$\left(\frac{m}{\gamma}\right)^2 \int_0^t dt' \int_0^t dt'' \left\{ \frac{(3\pi a^2)^2}{m^2} \langle \Delta\eta_y(t')\Delta\eta_y(t'')\rangle \langle \Delta\omega_z(t')\Delta\omega_z(t'')\rangle \right\}$$

$$= \frac{(3\pi a^2)^2}{\gamma^2} \langle \Delta\eta(t)^2 \rangle \int_0^t dt' \int_0^t dt'' \frac{k_B T}{I} e^{-\frac{\gamma'}{I}|t'-t''|}$$

$$= \frac{(3\pi a^2)^2}{\gamma^2} \langle \Delta\eta(t)^2 \rangle \frac{k_B T}{I} \int_0^t dt' \left\{ \int_0^{t'} dt'' \, e^{-\frac{\gamma'}{I}(t'-t'')} + \int_{t'}^t dt'' \, e^{-\frac{\gamma'}{I}(t''-t')} \right\}$$

$$= \frac{(3\pi a^2)^2}{\gamma^2} \langle \Delta\eta(t)^2 \rangle \frac{2k_B T}{I} \left\{ \frac{I}{\gamma'}\left(t - \frac{I}{\gamma'}\right) + \left(\frac{I}{\gamma'}\right)^2 e^{-\frac{\gamma'}{I}t} \right\}.$$

The much longer observed time can adapt with the approximation of $t - \frac{I}{\gamma'} \cong t, e^{-\frac{\gamma'}{I}t} \cong 0$.

$$\cong \frac{(3\pi a^2)^2}{\gamma^2} \langle \Delta\eta(t)^2 \rangle \frac{2k_B T}{I} \frac{I}{\gamma'} t$$

$$= \frac{3}{8} \frac{k_B T}{\gamma} \frac{\langle \Delta\eta(t)^2 \rangle}{\eta^2} \times t$$

Another term:

$$\left(\frac{m}{\gamma}\right)^2 \int_0^t dt' \int_0^t dt'' \frac{\langle R(t')R(t'') \rangle}{m^2} = \frac{1}{\gamma^2} \int_0^t dt' \int_0^t dt'' \, 2B\delta(t' - t'')$$

$$= \frac{2B}{\gamma^2} t = \frac{2k_B T}{\gamma} \left\{ 1 - \frac{15}{52} \frac{\langle \Delta\eta(t)^2 \rangle}{\eta^2} \right\} t.$$

Combining each term gives us:

$$\langle \Delta x(t)^2 \rangle = \frac{3}{8} \frac{k_B T}{\gamma} \frac{\langle \Delta\eta(t)^2 \rangle}{\eta^2} t + \frac{3}{8} \frac{k_B T}{\gamma} \frac{\langle \Delta\eta(t)^2 \rangle}{\eta^2} t + \frac{2k_B T}{\gamma} \left\{ 1 - \frac{15}{52} \frac{\langle \Delta\eta(t)^2 \rangle}{\eta^2} \right\} t$$

$$= \frac{2k_B T}{\gamma} \left\{ 1 + \frac{9}{104} \frac{\langle \Delta\eta(t)^2 \rangle}{\eta^2} \right\} \times t. \tag{S22}$$

Eq. (S22) demonstrates that the memory effect of Brownian rotation results in a larger displacement in the MSD. A maximum discrepancy of 5% was observed for the water–ethanol mixture. To account for this discrepancy, the inhomogeneous parameter can be estimated as approximately

$$\frac{\sqrt{\langle \Delta\eta^2 \rangle}}{\eta} \sim 0.76.$$

This discrepancy is attributed to a threefold increase in viscosity due to the mixing of water and ethanol, underscoring the inhomogeneous properties of the solution as a key factor influencing Brownian motion.

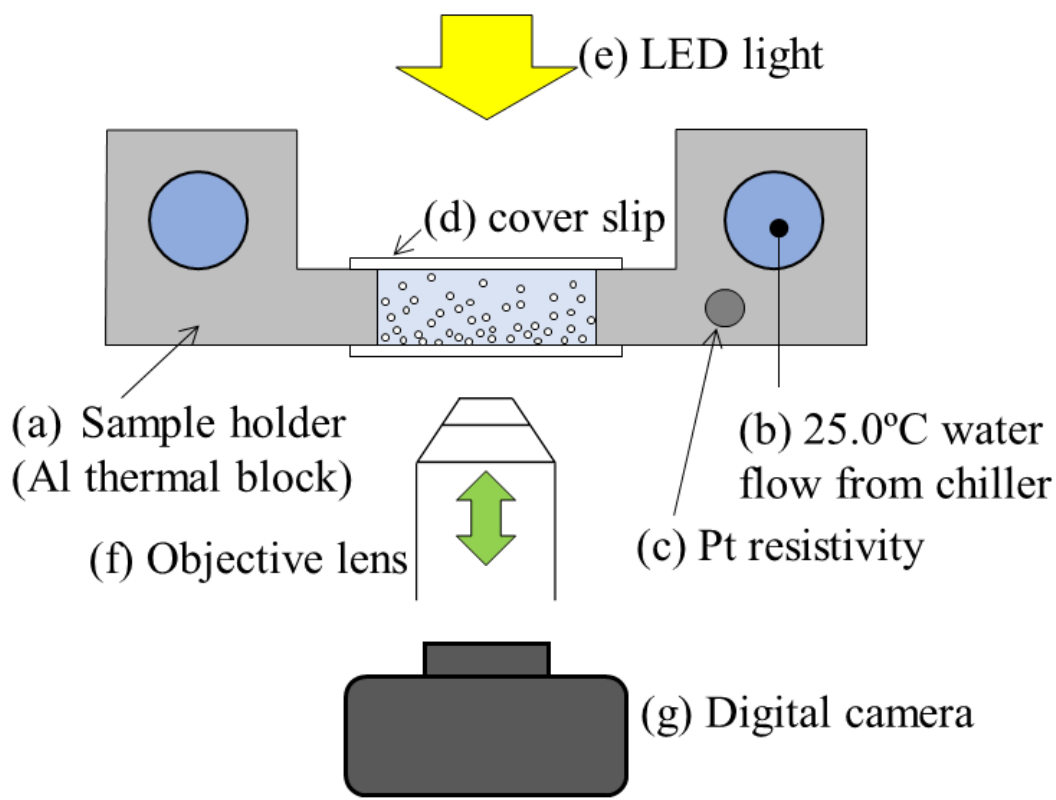

FIG. S1. Schematic of the experimental setup. The sample holder included (a) an aluminum thermal block, (b) a 25.0 °C liquid flow from a chiller, and (c) temperature control via Pt resistivity measurements. Suspensions of polystyrene beads in water–ethanol mixtures were sealed with (d) cover slips and silicon grease. Videos and photographs were captured using an inverted optical microscope, with (e) an LED light source, (f) a movable objective lens, and (g) a digital camera.

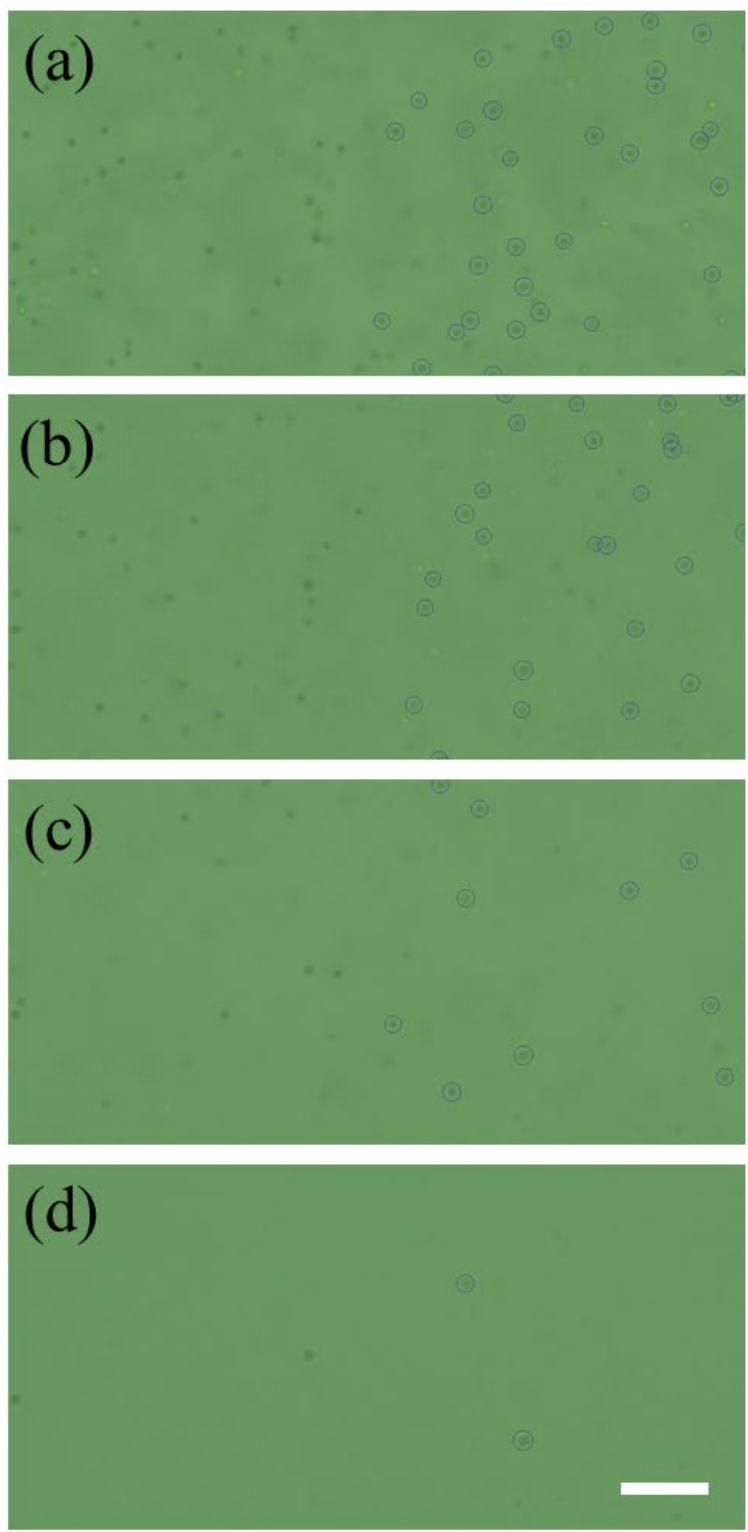

FIG. S2. Photographs acquired after 180 min of sedimentation, displayed in 2000 × 1000-pixel sections at heights of (a) 66.626, (b) 399.756, (c) 799.512, and (d) 1066.016 μm. On the right side of each picture, circles indicating auto-detected particles have been overlaid. The scale bar indicates 30 μm.

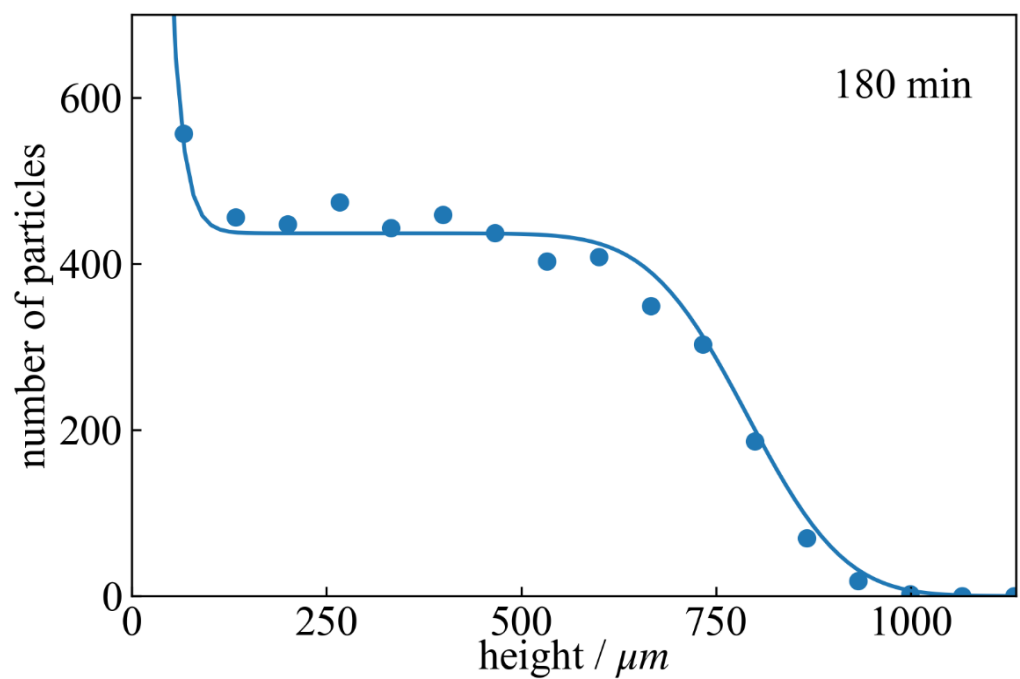

FIG. S3. Number of detected particles as a function of height before normalization (points). The sedimentation process is modeled using the Fokker–Planck equation (line).

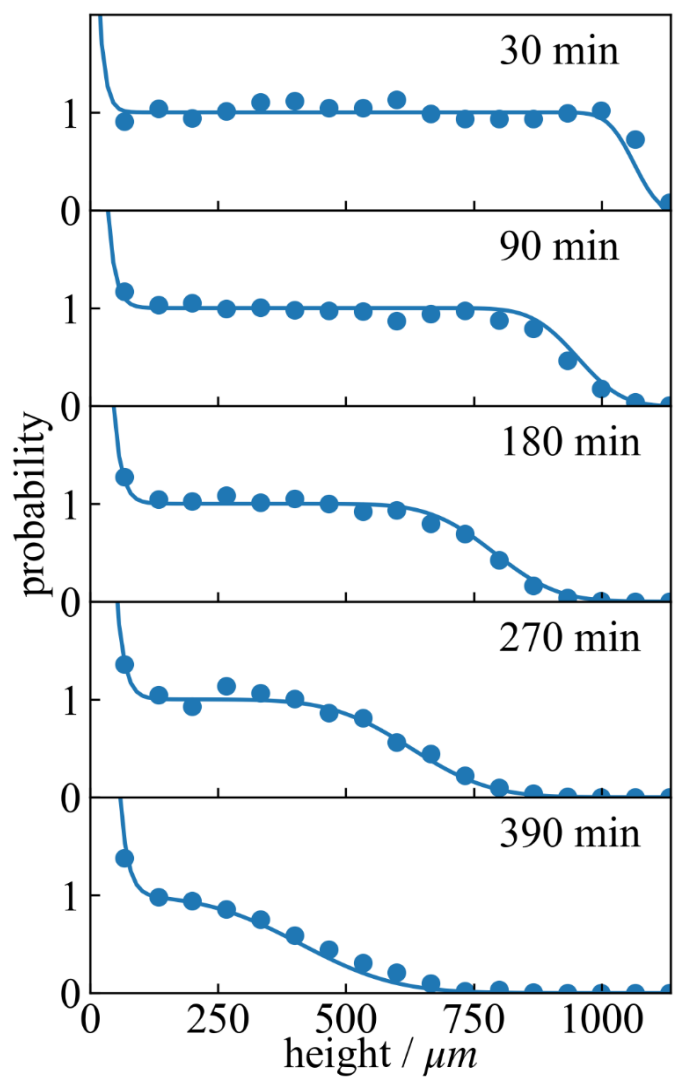

FIG. S4. Series of plots from the sedimentation process (falling-ball experiment) for pure water at 25.0 °C. The liquid viscosity value is obtained by optimizing the parameters to fit the Fokker–Planck equation for all the data, and the result shows excellent agreement with previously reported values for pure water (within 1% accuracy).